\documentclass[onecolumn,showpacs,preprintnumbers]{revtex4}
\usepackage{pifont}
\usepackage{subfigure}
\usepackage{epsfig}
\usepackage{CJK}
\usepackage{amsmath}
\usepackage{amsfonts}
\usepackage{amssymb}
\usepackage{color}
\usepackage{graphicx}
\usepackage{mathrsfs}
\usepackage[normalem]{ulem}
\UseRawInputEncoding
\def\be{\begin{equation}}
\def\ee{\end{equation}}
\def\ba{\begin{eqnarray}}
\def\ea{\end{eqnarray}}

\definecolor{dyellow}{rgb}{1.,0.8,.0}
\definecolor{myblue}{rgb}{.1,.1,.7}
\definecolor{dcyan}{rgb}{.0,.6,.6}
\definecolor{dmagenta}{rgb}{0.6,0.0,0.6}
\definecolor{brown}{rgb}{0.6,0.2,0.}
\definecolor{darkblue}{rgb}{.0,.0,0.5}
\definecolor{darkred}{rgb}{0.75,0.0,0.0}
\definecolor{orange}{rgb}{1.,.6,.0}
\definecolor{dorange}{rgb}{0.8,.4,.0}
\definecolor{darkgreen}{rgb}{0.0,0.6,0.0}
\definecolor{purple}{rgb}{.4,.0,.4}
\definecolor{lightgrey}{rgb}{0.7, 0.7, 0.7}
\definecolor{grey}{rgb}{0.4, 0.4, 0.4}

\newcommand{\bea}{\begin{eqnarray}}
\newcommand{\eea}{\end{eqnarray}}

\newcommand\btd{\raise 2pt
\hbox{$\hat\bigtriangledown$}\hskip 1.5pt}
\newcommand\bt{\raise 2pt
\hbox{$\bigtriangledown$}\hskip 1.5pt}

\newcommand{\omits}[1]{}

\begin{document}

\title{ Thermodynamic characteristics of two horizons coexistence region in 4D-EGB spacetime }
\author{Fang Liu$^{a,b}$, Yun-Zhi Du$^{a,b}$, Jian-Xin Sun$^{a,b}$, Huai-Fan Li$^{b,c}$}
\thanks{Corresponding author.\\ E-mail address: fangliuphys@sxdtdx.edu.cn (Fang Liu), duyzh13@lzu.edu.cn (Yun-Zhi Du), jxsun0128@163.com (Jian-Xin Sun), huaifan999@163.com (Huai-Fan Li)}

\affiliation{{\footnotesize $^a$ Department of Physics, Shanxi Datong University, Datong 037009, China}\\
{\footnotesize $^b$ Insitute of Theoretical Physics, Shanxi Datong University, Datong 037009,  China}\\
{\footnotesize $^c$ College of General Education, Shanxi College of Technology, Shuozhou 036000, China}}

\begin{abstract}
This paper investigates the thermodynamic properties of the coexistence region of two horizons in the charged 4-dimensional Einstein-Gauss-Bonnet (4D-EGB) spacetime. Initially, we apply the universal first law of thermodynamics to derive the corresponding thermodynamic quantities for the coexistence region between the black hole event horizon and the cosmological event horizon, subject to the relevant boundary conditions. Next we examine the thermal properties of the thermodynamic system described by these equivalent quantities. Our analysis reveals that the peak of the heat capacity as a function of temperature exhibits characteristics similar to those observed in a paramagnetic system under specific conditions. We further conclude that, under certain conditions, the heat capacity mirrors that of a two-level system formed by two horizons with distinct temperatures. By comparing the heat capacity of the 4D-EGB spacetime's equivalent thermodynamic system with that of a two-level system defined by the two horizons in the spacetime, we can estimate the number of microscopic degrees of freedom at the two horizons. This findings sheds light on the quantum properties of de Sitter (dS) spacetime with two horizon interfaces and offers a novel approach to exploring the quantum properties of black holes and dS spacetime.
\end{abstract}

\maketitle





\section{Introduction}

As is well known, the charged 4-Dimension Einstein Gauss-Bonnet (4D-EGB) black hole in de Sitter spacetime features both a black hole horizon and a cosmological horizon. The thermodynamic quantities associated with these horizons satisfy the first law of thermodynamics, and the corresponding entropies adhere to the area formulae\cite{DG-2020,PGSF-2022,AB-2021a,AB-2021b,MSC-2020,PGSF-2021,MG-2024}.  In recent years, the study of black hole properties in de sitter spacetime has garnered significant attention \cite{RGC-2002a,GAM-2021,AC-2024,PH-2024,HMS-2024,YPS-2024,GF-2024,AB-2024,HLZ-2024a,YS-2006,MU-2009,SM-2018,DK-2016,FS-2019,SH-2020,FS-2021,BPD-2013,SB-2016,Sl-2024,JM-2016,PK-2017,LCZ-2016,LCZ-2019,YBM-2020,JK-2024,VC-2023,PRA-2022,JD-2020,
   YBD-2023}.
Typically, the radiation temperatures of the two horizons are not equal. When the spacetime parameters are fixed within a certain range, research on the thermodynamic properties of dS spacetime, which features both a black hole event horizon and a cosmological event horizon, often treats these horizons as independent thermodynamic systems. However, the event horizons in dS spacetime are not truly independent; they are functions of the spacetime parameters. As the spacetime parameters vary, the thermodynamic quantities on the two horizons also change. Therefore, the thermodynamic quantities corresponding to the two horizons are interrelated, and it is not sufficient to consider them as independent systems. To properly study the thermodynamic properties of the coexistence region with two horizons, the first step is to identify a thermodynamic system that satisfies the boundary conditions for both horizons and captures the overall thermodynamic behaviors. we refer to this thermodynamic system as the equivalent thermodynamic system.

The idea of modifying General Relativity (GR) was first proposed by Wey \cite{HW-1918} and Eddington \cite{ASE-1921} in 1918, and the concept has persisted ever since. The current aims is to develop GR alternatives to address cosmological issues, such as cosmic acceleration in both the early and late universe. One such alternative theory is the Gauss-Bonnet gravity \cite{BJL-2007,GC-2006}.
The Gauss-Bonnet term, a higher-order curvature correction to the Einstein-Hilbert action, is particularly important in higher-dimensional theories and string theory. When incorporated into gravity, the Gauss-Bonnet term introduces non-trivial effects that do not arise in the standard Einstein-Hilbert framework.

However, gravitational dynamics in 4-dimensional spacetime are unaffected by the GB term. To address this limitation, D. Glavan and C. Lin recently proposed a novel 4-dimensional Einstein-Gauss-Bonnet (4D-EGB) gravity \cite{DG-2020}. This theory introduces the GB term with a factor $1/(d-4)$, thereby bypassing the stringent requirements of Lovelock＊s theorem \cite{DL-1971} and avoiding Ostrogradsky instability. In recent years, this new approach has become a valuable framework for exploring the physics and thermodynamics of black holes \cite{MSC-2020,PGSF-2021,MG-2024,GAM-2021,PGS-2020,FR-2024,MA-2024,SM-2024a,SM-2024b,SJY-2024,CM-2024,SWW-2020,JS-2024,YL-2023,YL-2024,RGC-2013}.

The first step in this analysis is to identify the equivalent thermodynamic quantities in the coexistence region between the black hole event horizon and the cosmic event horizon, ensuring they satisfy the first law of black hole thermodynamics. By examining the heat capacity of this coexistence region, described by these equivalent thermodynamic quantities, we find that the heat capacity curves of the spacetime resembles that of a two-level system, specifically following the Schottky specific heat curve. This transformation of the curve aligns with the heat capacity behavior observed when the two horizon interfaces are treated as a two-level system. This result deepens our understanding of 4D-EGB spacetime and opens new avenues for studying its thermodynamic properties. Additionally, it provides a novel approach for investigating the microscopic properties of particles inside black holes.

The structure of this paper is organized as follows: In the second section, we examine the conditions for the existence of black hole event horizon and cosmological event horizon in 4D-EGB spacetime, along with the effects of the parameters on the existence of these two horizons. The third section provides the equivalent thermodynamic quantities for 4D-EGB spacetime that satisfy the relevant boundary conditions. In the fourth section, we explore the heat capacity of spacetime as described by the these equivalent thermodynimic quantities. We discuss the transformation law of the heat capacity after fixing the space charge $Q$, cosmology constant $l$, and GB factor $\alpha$. Additionally, we compare the changes in the heat capacity of spacetime with equivalent temperature to those of a two-level systems. This paper concludes with a summary in the sixth section.  we use the units $G_d= \hbar=k_B=c=1$ in this paper.

\section{4D-EGB spacetime}
\label{two}
The Einstein-Gauss-Bonnet-de Sitter gravity in the $d$-dimensional spacetime is described by \cite{DK-2012,RGC-2002b}, which solves the Einstein-Maxwell field equation.
\begin{eqnarray}
I=\frac{1}{16\pi }\int{{{d}^{d}}}x\sqrt{g}[R-2\Lambda +\frac{\alpha }{d-4}({{R}_{\mu \nu \gamma \delta }}{{R}^{\mu \nu \gamma \delta }}-4{{R}_{\mu \nu }}{{R}^{\mu \nu }}+{{R}^{2}})-4\pi {{F}_{\mu \nu }}{{F}^{\mu \nu }}],\label{I}
\end{eqnarray}
where $g$ is the determinant of the metric ${{G}_{\mu \nu }}$, $R$ is the Ricci scalar and $\Lambda$ is the positive cosmological constant. The GB coupling $\alpha$ has dimension $[L]^2$ and can be identified with the inverse string tension with positive value if the theory is incorporated in string theory \cite{CHN-2018}, thus we shall consider only the case $\alpha>0$. $F_{\mu \nu }$ is the Maxwell field strength defined as $F_{\mu \nu }=\partial_{\mu} A_{\mu}-\partial_{\nu} A_{\nu}$ with vector potential $A_{\mu}$. The action admits a static black hole solution with metric
\begin{eqnarray}
d{{s}^{2}}=-f(r)d{{t}^{2}}+{{f}^{-1}}d{{r}^{2}}+{{r}^{2}}(d{{\theta }^{2}}+si{{n}^{2}}\theta d{{\varphi }^{2}}),\label{ds}
\end{eqnarray}
the metric function is
\begin{equation}
f(r)=k+\frac{{{r}^{2}}}{2\alpha }\left[ 1-\sqrt{1+4\alpha \left( \frac{2M}{{{r}^{3}}}-\frac{{{Q}^{2}}}{{{r}^{4}}}+\frac{1}{{{l}^{2}}} \right)} \right],  \label{fr}
\end{equation}
$M$ and $Q$ are the mass and charge of black hole, respectively. Without loss of the generality, we can set $k=1,0$ and $k=-1$, corresponding to spherical, flat and hyperbolic topologies of the black hole horizon, respectively.
The location of the black hole event horizon $r_+$ and the cosmic event horizon $r_c$, satisfy the relation $f(r_+,r_c)=0$. The mass $M$ can be expressed from Eq.(\ref{fr}) as
\begin{equation}
M=\frac{{{r}_{+,c}}}{2}\left( k+\frac{\alpha {{k}^{2}}}{r_{+,c}^{2}}+\frac{{{Q}^{2}}}{r_{+,c}^{2}}-\frac{r_{+,c}^{2}}{{{l}^{2}}} \right).\label{M}
\end{equation}
If $x=\frac{r_+}{r_c}$, $M$ can be written as
\begin{eqnarray}
&M&=\frac{k{{r}_{c}}x(1+x)}{2(1+x+{{x}^{2}})}+\frac{\alpha {{k}^{2}}(1+x)(1+{{x}^{2}})}{2{{r}_{c}}x(1+x+{{x}^{2}})}+\frac{{{Q}^{2}}(1+x)(1+{{x}^{2}})}{2{{r}_{c}}x(1+x+{{x}^{2}})},\nonumber\\
&\frac{1}{{{l}^{2}}}&=\frac{k}{r_{c}^{2}(1+x+{{x}^{2}})}-\frac{\alpha {{k}^{2}}}{r_{c}^{4}x(1+x+{{x}^{2}})}-\frac{{{Q}^{2}}}{r_{c}^{4}x(1+x+{{x}^{2}})}.\label{l2}
\end{eqnarray}
When $l^2=12\beta ({{Q}^{2}}+\alpha {{k}^{2}})$,($\beta \geq1$ is a constant.) we obtain
\begin{eqnarray}
r_{c}^{2}=\frac{6\beta k({{Q}^{2}}+\alpha {{k}^{2}})}{(1+x+{{x}^{2}})}\left( 1\pm \sqrt{1-\frac{(1+x+{{x}^{2}})}{3x\beta {{k}^{2}}}} \right),\nonumber\\
\frac{({{Q}^{2}}+\alpha {{k}^{2}})}{r_{c}^{2}}=\frac{xk}{2}\left( 1-\sqrt{1-\frac{(1+x+{{x}^{2}})}{3x{{k}^{2}}\beta }} \right).\label{rc21}
\end{eqnarray}
The radiation temperature on the two horizons can be expressed as
\begin{eqnarray}
&{{T}_{+}}&=\frac{f'({{r}_{+}})}{4\pi }=\frac{(1-x)}{4\pi (2\alpha k+r_{c}^{2}{{x}^{2}})(1+x+{{x}^{2}})}\left( k{{r}_{c}}x(1+2x)-\frac{({{Q}^{2}}+\alpha {{k}^{2}})(1+2x+3{{x}^{2}})}{{{r}_{c}}x} \right),\nonumber\\
&{{T}_{c}}&=-\frac{f'({{r}_{c}})}{4\pi }=\frac{(1-x)}{4\pi (1+x+{{x}^{2}})(2\alpha k+r_{c}^{2})}\left( k{{r}_{c}}(2+x)-\frac{({{Q}^{2}}+\alpha {{k}^{2}})(3+2x+{{x}^{2}})}{{{r}_{c}}x} \right) .\label{Tc}
\end{eqnarray}
The entropy of the two horizons is calculated according to the formula $S=\int{T^{-1}\frac{\partial M}{\partial r}}dr$ \cite{RGC-2002b,CHN-2018,HLZ-2024b} as
\begin{eqnarray}
&{{S}_{+}}&=\pi r_{+}^{2}+4\pi \alpha k\ln {{r}_{+}}=\pi r_{+}^{2}+2\pi \alpha k\ln \frac{{{A}_{+}}}{{{A}_{0}}},\nonumber\\
&{{S}_{c}}&=\pi r_{c}^{2}+4\pi \alpha k\ln {{r}_{c}}=\pi r_{c}^{2}+2\pi \alpha k\ln \frac{{{A}_{c}}}{{{A}_{0}}},\label{S}
\end{eqnarray}
in these expressions, ${A}_{0}$ is a constant with units of area.

\begin{figure*}[htb]
\subfigure[$~k=1,{{Q}^{2}}+\alpha{{k}^{2}}=1,l^2=70$]{\includegraphics[width=4.8cm,height=4.cm]{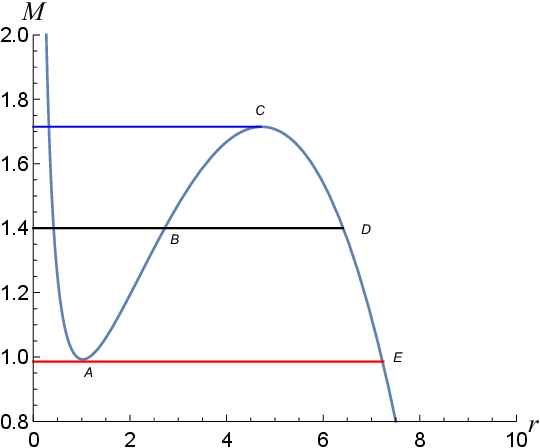}}~~
\subfigure[$~k=1,l^2=70$]{\includegraphics[width=4.8cm,height=4.cm]{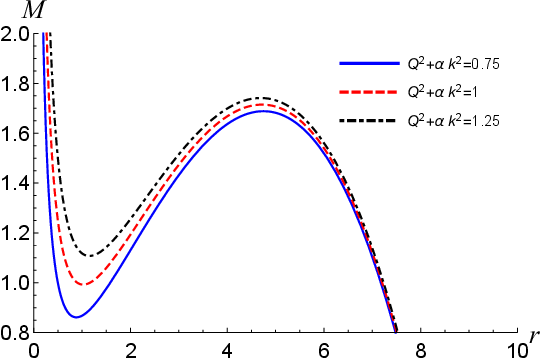}}~~
\subfigure[$~k=1,{{Q}^{2}}+\alpha {{k}^{2}}=1$]{\includegraphics[width=4.8cm,height=4.cm]{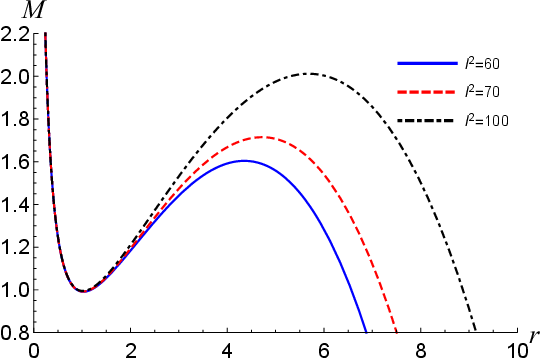}}
\vskip -1mm \caption{The behavior of $M$ as a function of $r$ for different values of $({{Q}^{2}}+\alpha {{k}^{2}})$ and $\l^2$ when $k=1$}\label{Fig.mr}
\end{figure*}

We plot the mass $M$ as a function of $r$ in FIG. \ref{Fig.mr}.
We can see from FIG. \ref{Fig.mr} (a), when $k=1,{{Q}^{2}}+\alpha {{k}^{2}}=1,l^2=70$, the energy of spacetime satisfied $M_A\leq M\leq M_C$, the 4D-EGB spacetime has an inner black hole horizon $r_-$, a black hole horizon $r_+$ and a cosmological horizon $r_c$.  At the local maximum, denoted as point C, the black hole event horizon and the cosmological event horizon coincide, corresponding to the massive energy state of the black hole event horizon and the cosmological event horizon in spacetime, where $M=M_C$. At the local minimum, point A, the inner and outer horizons of the black hole coincide, representing the small energy state of the black hole horizon and the cosmic horizon in spacetime, where $M=M_C$. When $M>M_C$ and $M<M_A$, no black holes exist in 4D-EGB spacetime \cite{CVJ-2020}.

At the local minimum point A and the local maximum point C, the black hole event horizon radius satisfies extremum condition
\begin{equation}
\frac{\partial M}{\partial r}=k-\frac{{{Q}^{2}}+\alpha {{k}^{2}}}{{{r}^{2}}}-\frac{3{{r}^{2}}}{{{l}^{2}}}=0.\label{pmr}
\end{equation}
By solving the equation above, we can get
\begin{equation}
r_{C,A}^{2}=\frac{{{l}^{2}}}{6}\left( k\pm \sqrt{{{k}^{2}}-\frac{12({{Q}^{2}}+\alpha {{k}^{2}})}{{{l}^{2}}}} \right).\label{r}
\end{equation}
Given the spacetime parameters $k$, ${{Q}^{2}}+\alpha {{k}^{2}}$, $l^2$ , substituting Eq. (\ref{r}) into Eq. (\ref{l2}) yeilds the maximum energy $M_C$ and minimum energy $M_A$ corresponding to the two horizon coexistence region as follows
\begin{eqnarray}
&{{M}_{c}}&=\frac{l}{3\sqrt{6}}{{\left( k+\sqrt{{{k}^{2}}-\frac{12({{Q}^{2}}+\alpha {{k}^{2}})}{{{l}^{2}}}} \right)}^{1/2}}\left( 2k-\sqrt{{{k}^{2}}-\frac{12({{Q}^{2}}+\alpha {{k}^{2}})}{{{l}^{2}}}} \right),\nonumber\\
&{{M}_{A}}&=\frac{l}{3\sqrt{6}}{{\left( k-\sqrt{{{k}^{2}}-\frac{12({{Q}^{2}}+\alpha {{k}^{2}})}{{{l}^{2}}}} \right)}^{1/2}}\left( 2k+\sqrt{{{k}^{2}}-\frac{12({{Q}^{2}}+\alpha {{k}^{2}})}{{{l}^{2}}}} \right).\label{mac}
\end{eqnarray}
When the location of black hole event horizon satisfies
\begin{equation}
\frac{({{Q}^{2}}+\alpha {{k}^{2}})}{r_{c}^{2}{{x}^{2}}}=\frac{k(1+2x)}{(1+2x+3{{x}^{2}})},\label{rc2}
\end{equation}
we know that the radiation temperature at the black hole event horizon $T_+=0$ from Eq. (\ref{Tc}). At point C, the black hole horizon and the cosmological horizon coincide $r_+=r_c$, the radiation temperature of both horizons is zero. At point A, the radiation temperature on black hole event horizon is zero, the cosmological horizon corresponds to point A is point E. For the equal energy $M_A$, the radiation temperature on black hole cosmic horizon is not zero. It can be obtained by substituting Eq. (\ref{rc2}) into Eq. (\ref{Tc}).
\begin{equation}
{{T}_{c}}({{T}_{+}}=0)={{T}_{E}}=\frac{k{{r}_{c}}{{(1-x)}^{2}}(1+x)}{2\pi (2\alpha k+r_{c}^{2})(1+2x+3{{x}^{2}})}.\label{Te}
\end{equation}

From FIG. \ref{Fig.mr} (a), we can see that when ${{Q}^{2}}+\alpha {{k}^{2}}$ and the cosmological constant $l$ are fixed, the energy of the spacetime satisfies $M_A\leq M\leq M_C$, which represents the coexistence zone of the black hole event horizon and the cosmological event horizon. For points B and D in the figure, the corresponding values of energy $M$, ${{Q}^{2}}+\alpha {{k}^{2}}$ and the cosmological constant $l$ are the same. However, in general, the radiation temperature $T_+$ at the black hole event horizon (Point B) is not equal to the temperature $T_c$ at the cosmological event horizon (Point D). In this region, two thermodynamic subsystems with different temperatures exist, corresponding to the black hole event horizon and the cosmological event horizon, respectively. Therefore, spacetime in this region is stable but in a non-thermal equilibrium state. At point A, where the black hole event horizon and inner event horizon coincide, the radiation temperature of the black hole event horizon is zero, corresponding to a cold black hole. This point also marks the endpoint of spacetime with two different radiation temperatures coexisting. For the same parameters, point E represents the transition point where spacetime shifts from two thermodynamic subsystems to single thermodynamic system with only the cosmological horizon. This is also the endpoint for spacetime with two thermodynamic subsystems. Point C, where $r_+=r_c$, is called the Nariai black hole. It is the point where the Hawking radiation temperature of both horizons becomes zero, and it marks the starting point of spacetime with two different temperature subsystems.

Thus, the 4D-EGB spacetime exhibits distinct thermodynamic properties in different regions, particularly in the coexistence region of two different temperature $T_+$ and $T_c$ . This coexitence region is one of the theoretical topics of great interest \cite{RGC-2002a,GAM-2021,HLZ-2024a,YS-2006,MU-2009,SM-2018,DK-2016}. In recent years, researchers have explored the thermodynamic characteristics of this region by constructing various models and derived meaningful and thought-provoking conclusions \cite{FS-2019,SH-2020,FS-2021,BPD-2013,SB-2016,Sl-2024,JM-2016,PK-2017,LCZ-2016,LCZ-2019,YBM-2020}. Building on these studies of the thermodynamic properties of dS spacetime, we use new constraints \cite{HLZ-2024b,CVJ-2020} to identify the equivalent thermodynamic quantities for the coexistence region of two horizons that satisfy the first law of black hole thermodynamics and the boundary conditions.
\section{The equivalent thermodynamic quantity of two horizons in 4D-EGB spacetime}
\label{three}
Based on the analysis in the previous section, we need to establish a thermodynamic system that characterizes the global properties of the region of spacetime with two subsystems. This system must satisfy specific boundary conditions and be universally applicable. By considering the relationship between the black hole horizon and the cosmological horizon, we derive the effective thermodynamic quantities and the corresponding first law of black hole thermodynamics \cite{LCZ-2016,LCZ-2019}.
\begin{equation}
dM={{T}_{eff}}dS-{{P}_{eff}}dV+{{\phi }_{eff}}dQ, \label{dm}
\end{equation}
here the thermodynamic volume is that between the black hole horizon and the cosmological horizon, namely \cite{DK-2016,FS-2019,SH-2020}
\begin{equation}
V={{V}_{c}}-{{V}_{+}}=\frac{4\pi }{3}r_{c}^{3}(1-{{x}^{3}}). \label{V}
\end{equation}
The equivalent entropy of thermodynamic system can be written as
\begin{equation}
S=\pi r_{c}^{2}\left( f(x)+\frac{4\alpha k}{r_{c}^{2}}\left( \ln \frac{r_{c}^{2}}{{{A}_{0}}}+\ln {{f}_{1}}(x) \right) \right).
\label{S1}
\end{equation}
In this expression,
\begin{equation}
f(x)=\frac{8}{5}{{(1-{{x}^{3}})}^{2/3}}+\frac{2}{5(1-{{x}^{3}})}-1,~~~
\ln {{f}_{1}}(x)=\ln (1-{{x}^{3}})+\frac{3-2{{x}^{3}}}{3(1-{{x}^{3}})}.\label{fx}
\end{equation}
When Eq. (\ref{rc2}) is satisfied, the equivalent temperature of spacetime should be the cosmic event horizon radiation temperature $T_E$.
According to the above, we obtain the equivalent temperature $T_{eff}$, equivalent pressure $P_{eff}$ and equivalent potential $\phi_{eff}$ as following
\begin{eqnarray}
{{T}_{eff}}&=&\frac{{{r}_{c}}(1-x)}{4\pi {{x}^{5}}(2\alpha k+r_{c}^{2})}\left[ k((1+x)(1+{{x}^{3}})-2{{x}^{2}})-\frac{({{Q}^{2}}+\alpha {{k}^{2}})[(1+x+{{x}^{2}})(1+{{x}^{4}})-2{{x}^{3}}]}{r_{c}^{2}{{x}^{2}}} \right],\nonumber\\
{{P}_{eff}}&=&-\frac{(1-x)}{16\pi {{x}^{5}}(2\alpha k+r_{c}^{2})}\left[ x(1+x)\left( f'(x)+\frac{4\alpha k{{f}_{1}}'(x)}{r_{c}^{2}{{f}_{1}}(x)} \right)\left( k-\frac{({{Q}^{2}}+\alpha {{k}^{2}})(1+{{x}^{2}})}{r_{c}^{2}{{x}^{2}}} \right) \right)\nonumber\\&-&\frac{2}{(1+x+{{x}^{2}})}\left( f(x)+\frac{4\alpha k}{r_{c}^{2}} \right)\left. \left( k(1+2x)-\frac{({{Q}^{2}}+\alpha {{k}^{2}})(1+2x+3{{x}^{2}})}{r_{c}^{2}{{x}^{2}}} \right) \right],\nonumber\\
{{\phi }_{eff}}&=&{{\left( \frac{\partial M}{\partial Q} \right)}_{V,S,\alpha }}=\frac{Q(1+x)(1+{{x}^{2}})}{{{r}_{c}}x(1+x+{{x}^{2}})}. \label{Teff}
\end{eqnarray}

\section{Heat capacity of the spacetime}

In thermodynamics, one of the primary concerns is how the thermodynamic quantities of achange in response to variations in its  external environment--specifically, the system's behavior under different conditions. In this section, we discuss the heat capacity in the coexistence region of the two horizons. we demonstrate that the thermodynamic quantities of the two-horizon coexistence region are determined when the parameters $k$, $Q$, $\alpha$ and $l$ are fixed. Therefore, we examine the heat capacity for specific values of these parameters.

The heat capacity of the system can be described as
\begin{equation}
{{C}_{Q,l\text{,}\alpha }}={{T}_{eff}}{{\left( \frac{\partial S}{\partial {{T}_{eff}}} \right)}_{Q,l\text{,}\alpha }}={{T}_{eff}}{{\left( \frac{dS/dx}{d{{T}_{eff}}/dx} \right)}_{Q,l\text{,}\alpha }}.\label{cqla}
\end{equation}
Notice the expression for the entropy of the system,
\begin{equation}
S=\pi r_{c}^{2}f(x)+4\alpha \pi k\left( \ln \frac{r_{c}^{2}}{{{A}_{0}}}+\ln {{f}_{1}}(x) \right)={{S}_{1}}+{{S}_{2}} ,
\label{S2}
\end{equation}
we know that the heat capacity of the system is composed of two parts,
\begin{equation}
{{C}_{Q,l,\alpha }}=C_{Q,l,\alpha }^{1}+C_{Q,l,\alpha }^{2},\label{cqls}
\end{equation}
here $C_{Q,l,\alpha }^{1}$ corresponding to the entropy $S_1=\pi r_{c}^{2}f(x)$, $C_{Q,l,\alpha }^{2}$ corresponding to the entropy $S_2=4\alpha \pi k\left( \ln \frac{r_{c}^{2}}{{{A}_{0}}}+\ln {{f}_{1}}(x) \right)$.
Substituting Eq. (\ref{rc21}) into Eq. (\ref{cqla}), we plot the heat capacity curve when $k=1$, $\alpha=0.4$ and ${{Q}^{2}}+\alpha {{k}^{2}}=1$.
\begin{figure*}[htb]
\subfigure[$~ k=1,\alpha=0.4.$]{\includegraphics[width=4.8cm,height=4.cm]{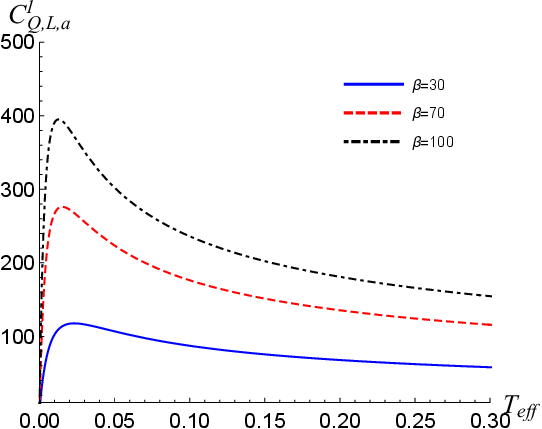}}~~
\subfigure[$~ k=1,\alpha=0.4.$]{\includegraphics[width=4.8cm,height=4.cm]{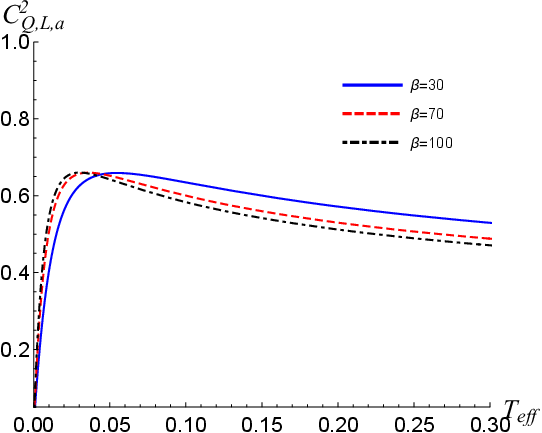}}~~
\subfigure[$~ k=1,\alpha=0.4.$]{\includegraphics[width=4.8cm,height=4.cm]{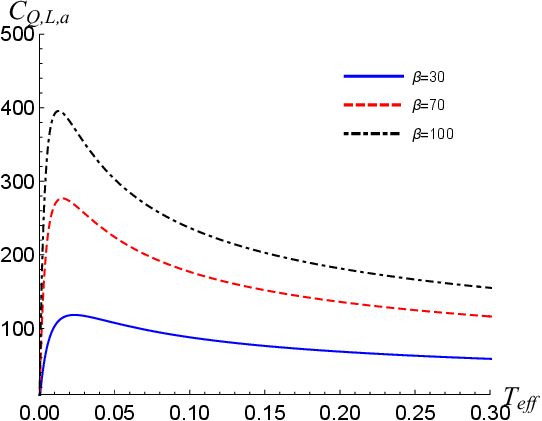}}
\vskip -1mm \caption{The behavior of the heat capacity as a function of ${T}_{eff}$ for different values of $\beta$.}\label{Fig.Cqa}
\end{figure*}
\begin{figure*}[htb]
\subfigure[$~ k=1,\beta=70.$]{\includegraphics[width=4.8cm,height=4.cm]{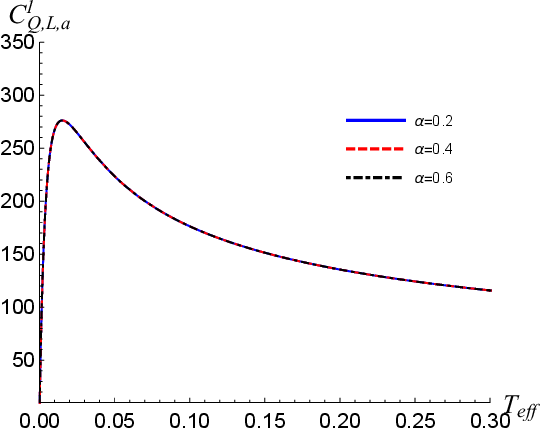}}~~
\subfigure[$~ k=1,\beta=70.$]{\includegraphics[width=4.8cm,height=4.cm]{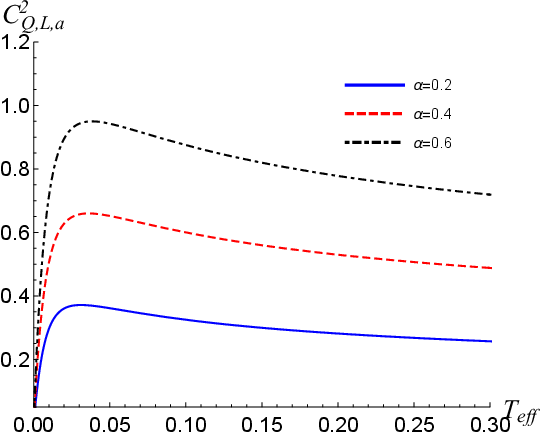}}~~
\subfigure[$~ k=1,\beta=70.$]{\includegraphics[width=4.8cm,height=4.cm]{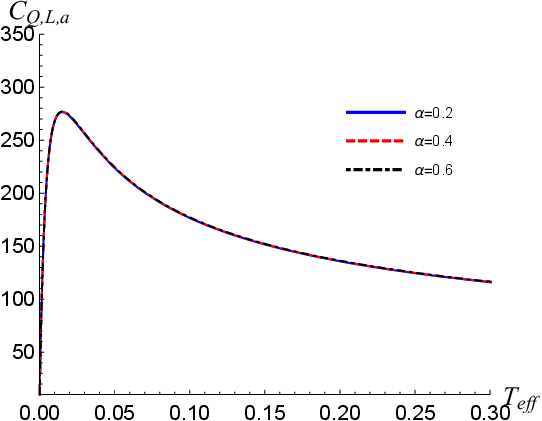}}
\vskip -1mm \caption{The behavior of the heat capacity as a function of ${T}_{eff}$ for different values of $\alpha$.}\label{Fig.Cqb}
\end{figure*}
\begin{figure*}[htb]
\subfigure[$~ k=1,\alpha=0.4.$]{\includegraphics[width=4.8cm,height=4.cm]{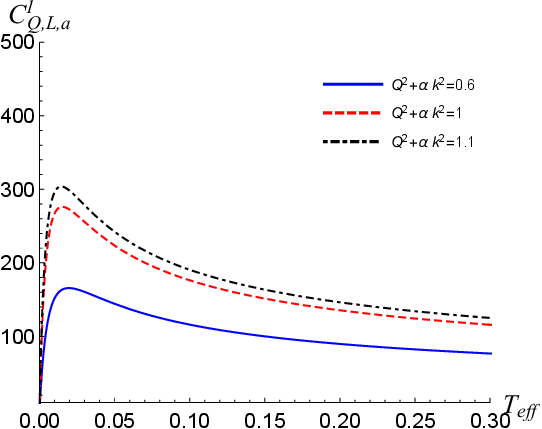}}~~
\subfigure[$~ k=1,\alpha=0.4.$]{\includegraphics[width=4.8cm,height=4.cm]{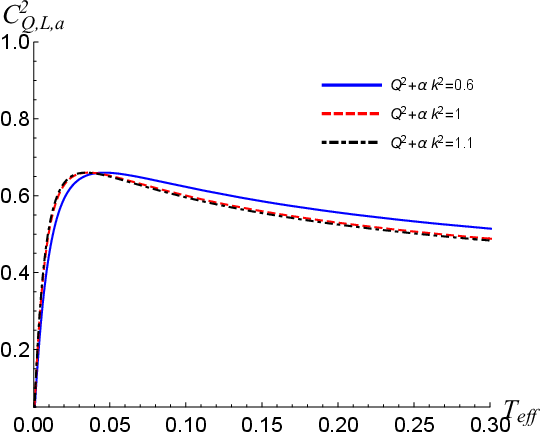}}~~
\subfigure[$~ k=1,\alpha=0.4.$]{\includegraphics[width=4.8cm,height=4.cm]{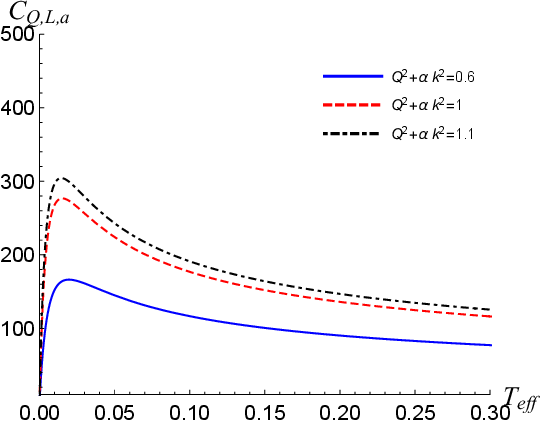}}
\vskip -1mm \caption{The behavior of the heat capacity as a function of ${T}_{eff}$ for different values of $({{Q}^{2}}+\alpha {{k}^{2}})$.}\label{Fig.Cqc}
\end{figure*}

The heat capacity of the system in the coexistence zone of two horizons varies with temperature under different conditions, as shown in FIG. \ref{Fig.Cqa}, \ref{Fig.Cqb} and \ref{Fig.Cqc}. The continuous change curve of heat capacity ${C}_{Q,l,\alpha }$ with the effective temperature $T_{eff}$ exhibits two distinct characteristics. First, the heat capacity curve reaches a maximum value at some temperature. Second, the heat capacity ${C}_{Q,l,\alpha }$ approaches zero as $T_{eff}$ tends to zero under all conditions. Both $C_{Q,l,\alpha }^{1}$ and $C_{Q,l,\alpha }^{2}$ exhibit the same features. It is well known that Gauss-Bonnet gravity enhances the ultraviolet (UV) behavior of the theory by introducing the quadratic Ricci curvature term $R^2$. From the $C_{Q,l,\alpha }^{1}-T_{eff}$ and $C_{Q,l,\alpha }^{2}-T_{eff}$ curves, it can be seen that the modified heat capacity curve $C_{Q,l,\alpha }^{2}-T_{eff}$ (due to the Gauss-Bonnet fcator $\alpha$) aligns with the curve $C_{Q,l,\alpha }^{,}-T_{eff}$ that lacks the modified term. These two characteristics of heat capacity $C_{Q,l,\alpha }$ reflect, in part, the quantum properties of the microscopic particles in the equivalent thermodynamic system. To further exploring the quantum properties of the internal microscopic particles in the heat capacity of the two-horizon coexistence region, we analyze the system as a two-level system.

\section{the two-energy-level system}
In a typical thermodynamic system, the atoms that make up the system posses a fixed magnetic moment ${\vec{\mu}}$ (or electric dipole moment $\vec E$ ). When such a substance is placed in an external magnetic field $\vec H$ (or electric field $\vec E$), the magnetic moment (or electric dipole moment $\vec P$) aligns with the direction of the field. As a result, the potential energy $-{\vec {\mu}}\cdot{\vec H}$ ($-{\vec P}\cdot{\vec E}$) is minimized, in accordance with the principle of minimum energy. For a system composing of such atoms, the partition function of a single particle, neglecting the interactions between the particles, is equivalent to that of a paramagnetic system, which can be modeled as a two-energy-level system.
\begin{equation}
Z(T,H,1)={{e}^{\beta \varepsilon }}+{{e}^{-\beta \varepsilon }}=2\cosh (\beta \varepsilon ).\label{Z}
\end{equation}
From Eq. (\ref{Z}), we obtain thermodynamic function as
\begin{eqnarray}
S(T,H,N)&=&Nk[\ln 2\cosh (\beta \varepsilon )-\beta \varepsilon \tanh (\beta \varepsilon )],\nonumber\\
U&=&F+TS=-Nk\tanh (\beta \varepsilon ),\nonumber\\
{{C}_{H}}&=&{{\left( \frac{\partial U}{\partial T} \right)}_{H}}=Nk{{(\beta \varepsilon )}^{2}}{{\cosh }^{-2}}(\beta \varepsilon ). \label{STHN}
\end{eqnarray}
For a thermodynamic system in which microscopic particles with electric dipole moments $\vec P$ are in an electric field, the above equation $\vec H$ can be replaced by $\vec E$. Heat capacity of a two-level system is
\begin{equation}
{{C}_{H(E)}}=Nk{{\left( \frac{\Delta }{T} \right)}^{2}}\frac{{{e}^{\tfrac{\Delta }{T}}}}{{{(1+{{e}^{\tfrac{\Delta }{T}}})}^{2}}},\label{Che}
\end{equation}
here $ \frac{1}{\beta \varepsilon}=\frac{kT}{\varepsilon } $, $N$ is the number of microscopic particles, $k$ is the Boltzmann constant ($k=1.381\times {10}^{-23}J\cdot K^{-1}$).

\begin{figure*}[htb]
{\includegraphics[width=4.8cm,height=4.cm]{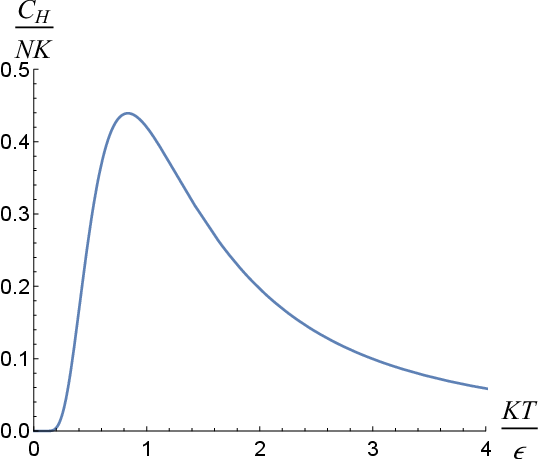}}~~
\caption{The behavior of $\frac{kT}{\varepsilon }- \frac{C_H}{Nk}$ .}\label{Fig.Ch}
\end{figure*}

We plot the $\frac{kT}{\varepsilon }- \frac{C_H}{Nk}$ curve in FIG. \ref{Fig.Ch}.
It can be shown that a two-level quantum system with a split energy spectrum follows the rule that $C\rightarrow0$ as $T\rightarrow0$, with the specific heat reaching a maximum value approximately when $\varepsilon \simeq kT$. FIG. \ref{Fig.Ch} illustrates the characteristics of a two-level system with an energy gap $\Delta =2 \varepsilon$, which is commonly known as
the Schorrky specific heat. One of its key features is that many physical systems can be effectively modeled as two-energy-level systems.

Since the 4D-EGB spacetime has two horizons with different temperatures, and these two horizons in a system with an equivalent temperature $T_{eff}$, the variation in heat capacity between the two horizons is a significant point of interest. We treat the two horizons as different energy levels within the equivalent temperature $T_{eff}$ and investigate the relationship between the heat capacity $C_{Q,l,\alpha }$ of the equivalent system and the heat capacity $\hat{C}_{Q,l,\alpha }$ of the two-energy-level system. This approach allows us to further explore the microscopic states of the particles in the system corresponding to the coexist regions of the two horizons. In this equivalent thermodynamic system, the difference between the two energy levels is
\begin{equation}
\Delta ={{\varepsilon }_{+}}-{{\varepsilon }_{c}}=\varepsilon ({{T}_{+}}-{{T}_{c}}),~T=T_{eff}/{\varepsilon}, \label{e}
\end{equation}
here $\varepsilon$ is a constant, when $\varepsilon=1$, (The value of $\varepsilon$ does not influence the transformation rule of the curve), substituting Eq. (\ref{e}) into Eq. (\ref{Che}), we obtain
\begin{equation}
{{\hat{C}}_{Q,l,\alpha }}=NK{{\left( \frac{\Delta }{{{T}_{eff}}} \right)}^{2}}\frac{{{e}^{\tfrac{\Delta }{{{T}_{eff}}}}}}{{{(1+{{e}^{\tfrac{\Delta }{{{T}_{eff}}}}})}^{2}}}, \label{hatc}
\end{equation}
\begin{equation}
\frac{\Delta }{{{T}_{eff}}}=\frac{{{x}^{5}}(1-{{x}^{2}})}{(1+x+{{x}^{2}})(2\alpha k+r_{c}^{2}{{x}^{2}})}\frac{\left( k[r_{c}^{2}x-4\alpha k]-\frac{({{Q}^{2}}+\alpha {{k}^{2}})[r_{c}^{2}{{(1+x)}^{2}}-4\alpha k]}{r_{c}^{2}x} \right)}{\left[ k((1+x)(1+{{x}^{3}}-2{{x}^{2}})-\frac{({{Q}^{2}}+\alpha {{k}^{2}})[(1+x+{{x}^{2}})(1+{{x}^{4}})-2{{x}^{3}}]}{r_{c}^{2}{{x}^{2}}} \right]}. \label{dt}
\end{equation}

We plot the variation in heat capacity curves when $NK=1$.
\begin{figure*}[htb]
\subfigure[$~k=1,~\alpha=0.4,~({{Q}^{2}}+\alpha {{k}^{2}})=1.$]{\includegraphics[width=4.8cm,height=4.cm]{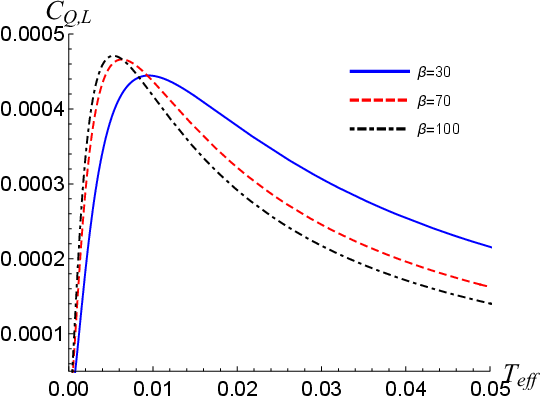}}~~
\subfigure[$~k=1,~\beta=70,~({{Q}^{2}}+\alpha {{k}^{2}})=1.$]{\includegraphics[width=4.8cm,height=4.cm]{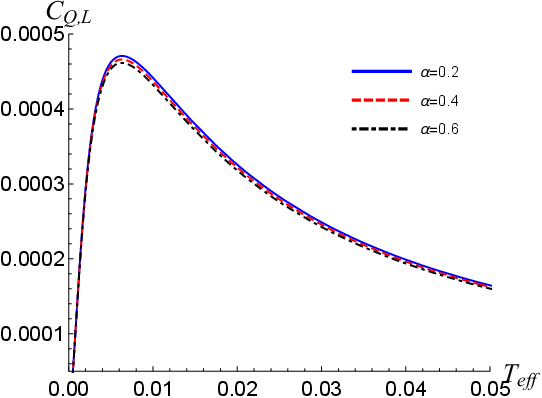}}~~
\subfigure[$~k=1,~\alpha=0.4,~\beta=70.$]{\includegraphics[width=4.8cm,height=4.cm]{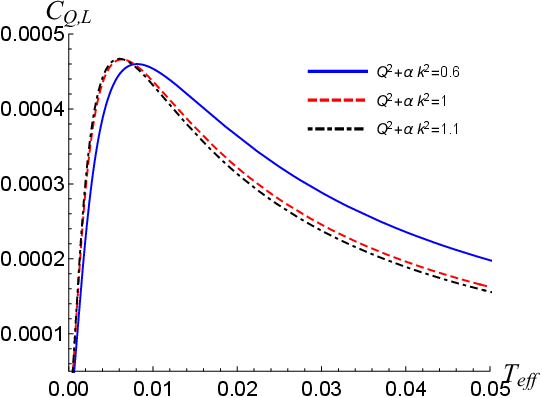}}
\vskip -1mm \caption{The behavior of the heat capacity as a function of ${T}_{eff}$ for different values of $~\alpha,~\beta,~({{Q}^{2}}+\alpha {{k}^{2}})$.}\label{Fig.Cqd}
\end{figure*}

When the horizons with different temperatures are considered as two distinct energy levels in an equivalent thermodynamic system, the microscopic components of the system at different horizons correspond to the particles at different energy levels. As shown in FIG. \ref{Fig.Cqc}, the continuous change curve of the heat capacity as a function of temperature in this system exhibits the typical characteristic of a two-energy-level system. First, the heat capacity curve reaches a maximum value as the temperature changes. Secon, the heat capacity ${\hat{C}}_{Q,l,\alpha }$ approaches zero as $T_{eff}\rightarrow0$. Using the method outlined in reference \cite{HLZ-2024a,HLZ-2024b}, and by comparing FIG. \ref{Fig.Cqa}, \ref{Fig.Cqb}, \ref{Fig.Cqc} with FIG. \ref{Fig.Ch},  we can approximate the number of microscopic particles between the two horizons in 4D-EGB spacetime.
\section{conclusion}
In this study, we model 4D-EGB spacetime as an equivalent thermodynamic system, deriving its thermodynamic quantity by considering the boundary conditions in de Sitte spacetime. Base on this framework, we present the equivalent thermodynamic quantity for the coexistence region of the two horizons and explor the heat capacity within this region. We find that the heat capacity of the spacetime, described by the equivalent thermodynamic quantity, exhibits the characteristics similar to the Schottky specific heat.

To explain the underlying cause of this behavior, we treat the black hole event horizon and the cosmological event horizon as two distinct energy levels within the equivalent thermodynamic system, thus modeling the system as a two-energy-level system. By analyzing the heat capacity of this two-level system, we observe that the heat capacity curve ${\hat{C}}_{Q,l,\alpha }-T_{eff}$ closely resembles the curve ${C}_{Q,l,\alpha }-T_{eff}$ of the equivalent thermodynamic system. Both $C_{Q,l,\alpha }^{1}$ and $C_{Q,l,\alpha }^{2}$ exhibit consistent behavior in terms of the change in heat capacity. All heat capacity curves-- ${\hat{C}}_{Q,l,\alpha }-T_{eff}$, ${C}_{Q,l,\alpha }-T_{eff}$, $C_{Q,l,\alpha }^{1}-T_{eff}$ and $C_{Q,l,\alpha }^{2}-T_{eff}$ follow the law ${\hat{C}}_{Q,l,\alpha }\rightarrow 0, {C}_{Q,l,\alpha }\rightarrow 0, C_{Q,l,\alpha }^{1}\rightarrow 0, C_{Q,l,\alpha }^{2}\rightarrow 0$ as $T_{eff}\rightarrow 0$. The heat capacity reaches a maximum value as the temperature changes.

The heat capacity of the 4D-EGB spacetime satisfies the general relation of two energy levels, as desceibed by Eq. (\ref{hatc}), corresponding to the famous Schottky specific heat. We also explor whether this heat capacity variation law is universal in dS spacetime, and how different parameters and spacetime dimensions influence the heat capacity curves of high-dimensional spacetime with multiple parameters. To answer this, further studies of various dS spacetime are necessary to identify a universal law.

Additionally, a more in-depth exploration of this subject will enhance our understanding of the quantum properties of black holes  and offer new insight into the thermodynamic properties of dS spacetime. Moving forward, the goal is to find a universal equivalent thermodynamic quantity for the two-horizon coexistence region that satisfies the boundary conditions, and investigate whether the thermodynamic system described by equivalent thermodynamics exhibits phenomena such as phase transition, critical point, and the Joule-Thomson effect.

\acknowledgments
This work was supported by the Natural Science Foundation of China (Grant No. 12375050, Grant No. 12075143),  the Natural Science Foundation of Shanxi Province (202303021211180), and the Doctoral Sustentation Fund of Shanxi Datong University.

\end{document}